\documentstyle[editedvolume,numreferences]{crckapb}

\newcommand{\be}{\begin{equation}}
\newcommand{\ee}{\end{equation}}
\newcommand{\rref}[1]{(\ref{#1})}

\begin{opening}
\title{INTERSECTION RULES AND OPEN BRANES}
\author{Riccardo Argurio}
\institute{Service de Physique Th\'eorique\\
Universit\'e Libre de Bruxelles, Campus Plaine, C.P.225\\
Boulevard du Triomphe, B-1050 Bruxelles, Belgium}
\end{opening}

\begin{document}

\begin{abstract}
A general rule determining how extremal branes can intersect in a 
configuration with zero binding energy is presented.
It is derived in a model independent way and without explicit use of
supersymmetry, solving a set of classical equations of motion. 
When specializing to M and type II 
theories, it is shown that some intersection rules can be consistently
interpreted as boundary rules for open branes ending on other branes. 
\end{abstract}

\section{Introduction}
Classical solutions of various supergravities are very interesting to
study in the context of string theories and M-theory, because they are 
often essential in establishing or corroborating the existence of
dualities relating (some compactified versions of) the above-mentioned
theories \cite{hulltown,witten,democracy}.
These $p$-brane solutions (for some reviews on $p$-branes, see
e.g. \cite{stelle,gauntlett,youm}) provide us with informations
on the long-range, low-energy fields produced by objects
which live in a more complete theory, i.e. a theory of 
superstrings or the `would-be' M-theory. Solutions involving 
several (classical) branes are thus useful in determining some
characteristics of the interactions between the quantum objects,
and in putting forward conjectures about the quantum dynamics of the
underlying theory.

The problem of studying the interactions between different branes
in string and M-theory will be addressed 
in this contribution considering supergravity
solutions which involve intersecting branes.
More precisely, we will be concerned with orthogonal intersections
of extremal branes. This means that each constituent brane saturates
a BPS bound, its mass being equal to its charge in the relevant units.
Considered on its own, such a single brane solution would preserve
half of the space-time supersymmetries. The intersecting brane
solutions that we will consider are such that the full solution
still preserves some (lower) fraction of supersymmetry. This is 
related to the fact that the binding energy of these configurations
vanishes. As we will show hereafter, these solutions are relevant to
the study of black hole physics, since they will provide, in the reduced
space-time, black holes with non-vanishing Bekenstein-Hawking entropy,
but which are nevertheless still supersymmetric and thus much more
tractable.

Historically, solutions allowing for a single $p$-brane were presented
most generally in \cite{horostrom}. The most important feature
of these solutions, in their extreme limit, is that they are characterized
entirely by a single harmonic function, which depends on
the coordinates transverse to the brane. 
In spring '96, Papadopoulos and Townsend \cite{papatown}
reinterpreted some 11 dimensional supergravity solutions found by 
G\"uven \cite{guven} as intersecting
branes and then used this interpretation to build new solutions. 
Soon after, Tseytlin further generalized in \cite{tseytlin_harm}
these solutions to include an independent harmonic function for each
(non-parallel) brane in the solution. The application of dualities 
to these particular solutions then predicted a lot of new configurations
involving all sorts of branes. For D-branes, these new solutions were
compatible with the supersymmetric intersections derived in string 
theory (see \cite{tasi,green_lect}).
The main common feature, besides the appearance
of the harmonic function associated to each brane, was the vanishing
of the binding energy.
The rule to build such solutions was simply to `superpose' the single brane 
solutions. This led to the formulation by Tseytlin of the `harmonic 
superposition rule' for orthogonally intersecting branes.
Still, the dimension of the intersection had to be determined case by
case, from supersymmetry arguments and/or by duality. The nature
of the argument strongly depended on the type of brane considered
(NS-, D- or M-brane).

The outline of the rest of my contribution is as follows. We will first
show how to derive the harmonic superposition rule from the equations
of motion of a general theory, which models supergravity. Supersymmetry
will not be an ingredient of this derivation, though it will (remotely)
motivate some of the ans\"atze made in order to solve the equations.
Almost as a byproduct, some of the equations of motion will reduce to 
a set of algebraic equations determining the dimension of the pairwise
intersections of the branes in the configurations (i.e. the `intersection
rules'). We will then proceed to the tentative deduction of some
brane dynamics from these solutions. In the case at hand we will consider
the possibility for some branes to open, with boundaries tied
to some other brane. For this to work, we have to check that the
charge of the open brane is still conserved. The mechanism by which
this is done sheds some light on the world-volume effective theory
of the brane on which the open brane ends. In the end we speculate
on the relevance of closed brane emission by other branes.
This talk is based on the two papers \cite{aeh,aehw}, written in collaboration
with F.~Englert, L.~Houart and P.~Windey.

\section{The Harmonic Superposition and the Intersection Rules}
In this section we will derive the harmonic superposition rule 
and the dimension of the intersection of extreme branes simply solving
a set of bosonic equations of motion, provided some particular
ans\"atze are made. 
A more detailed and step-by-step derivation can be found in \cite{aeh}.
As a starting point we take  a general action
in $D$ dimensions, which can be the bosonic part of a supergravity
action:
\be
I= \frac{1}{16\pi G_N^{(D)}} \int d^D x \sqrt{-g}
\left( R - \frac{1}{2} (\partial \phi)^2 -
\sum_I \frac{1}{2 n_I !}e^{a_I \phi}F_{n_I}^2 \right), \quad
I=1\dots {\cal M}. \label{action}
\ee
${\cal M}$ is the number of antisymmetric tensor fields, and 
we take $n_I\leq D/2$ for all field strengths. The metric
is written in the Einstein frame, and the coupling of the forms to
the dilaton in this frame is entirely governed by the constants $a_I$.
Note that since there is only one scalar, this theory is most suitable
to model 10 or 11 dimensional supergravities. Generalizations to include
several scalars can be found in the literature. Also, we did not include
for the moment Chern-Simons terms. A posteriori, they can be shown to
play no role in the determination of the classical solutions considered
here, but their presence will be crucial when we will discuss the
opening of the branes.

We begin now to simplify the problem taking the metric to be of a particular,
diagonal, form:
\be
ds^2=-B^2 dt^2+C_1^2dy_1^2+\dots+C_p^2dy_p^2+G^2dx_a dx_a, \qquad a=1\dots
D-p-1. \label{metric}
\ee
We have an $SO(D-p-1)$ symmetry left in the `overall transverse' space
of the $x$'s, which are taken to be the non-compact directions. Note
that there is no a priori $SO(p,1)$ symmetry, and that we will not
necessarily have a $p$-dimensional brane in the solution. The $y$'s
directions will be eventually compactified. Since the metric is
diagonal, we exclude for the moment solutions 
involving KK waves and monopoles. All functions in the problem depend
only on the $x_a$'s.

For the $n$-form field strengths, we have the choice between two
different ans\"atze:
\begin{eqnarray}
\mbox{Electric}& & F_{ty_1\dots y_{q_A}a}=\partial_a E_A, \label{electric}\\
\mbox{Magnetic}& & \tilde{F}_{ty_1\dots y_{q_A}a}=\partial_a E_A,
\label{magnetic}
\end{eqnarray}
where we have defined the dual field strength by:
\[ \tilde{F}_{\mu_1\dots\mu_{D-n}}=\sqrt{-g}e^{a\phi}
\epsilon_{\mu_1\dots\mu_D}F^{\mu_{D-n+1}\dots\mu_D}. \]
The space-time charges are thus respectively defined by:
\be
Q^{el}_A\sim \int *F_{q_A+2}, \qquad Q^{mag}_A\sim \int F_{D-q_A-2}.
\label{charges}\ee
$A=1\dots{\cal N}$, where ${\cal N}$ is the total number of
different (non-parallel) branes, electric and magnetic, in the solution.
This number can of course exceed the number of different $n$-forms.

We can now take the key steps which will enable us to solve quite
straightforwardly the equations of motion derived from the theory
above. These are the following two ans\"atze:
\begin{itemize}
\item Extremality, which (by experience) is enforced on the metric
by the condition:
\be BC_1\dots C_p G^{D-p-3}=1. \label{ansatz1} \ee
\item No-force condition between the constituent branes (in other words,
the requirement that the branes form a BPS marginal bound state). This is 
translated in our problem in the statement that to each brane is
associated one independent harmonic function, and that the solution
is completely characterized by these ${\cal N}$ harmonic functions.
\end{itemize}

These two conditions could in principle be found asking the solution to
preserve some supersymmetries, i.e. demanding that the equation
$\delta_{SUSY} \psi=0$ has non-trivial solutions. However this cannot
be done in this generic set up, i.e. for arbitrary $D$. 

The mathematical implementation of the 
second ansatz can be motivated as follows. We know that a single
brane solution is entirely determined by only one harmonic function. If
there are ${\cal N}$ branes in the configuration, 
but there is no binding energy, nothing prevents us from pulling
one of the branes apart from the others. Then the fields near that brane
should be a good approximation to the fields in the single brane solution.
Thus we see that we should expect exactly ${\cal N}$ independent functions
in the solution.\footnote{
See \cite{tseytlin_noforce} for a derivation of the intersection rules
based on the application of the no-force ansatz on the effective brane
actions, and see \cite{ohta} for a detailed discussion of the second
ansatz and its extension to non-extreme intersecting branes.}

These ${\cal N}$ independent functions are taken to be $H_A$ such that 
the $E_A$'s in \rref{electric} and \rref{magnetic} satisfy:
\be E_A \sim H_A^{-1}. \label{ansatz2} \ee
Then the equations of motion impose (see \cite{aeh} for the details)
$\partial_a \partial_a H_A=0$, which gives:
\be
H_A=1+\sum_k \frac{c_A Q_{A,k}}{|x^a-x^a_k|^{D-p-3}}. \label{multicenter}
\ee

Solving for the Einstein and the dilaton equations gives the following
metric and dilaton:
\be ds^2=-\prod_A H_A^{-2\frac{D-q_A-3}{\Delta_A}}dt^2+\sum_i
\prod_A H_A^{-2\frac{\delta^{(i)}_A}{\Delta_A}}dy_i^2+ 
\prod_A H_A^{2\frac{q_A+1}{\Delta_A}}dx_a dx_a, \label{metric_sol} \ee
\be e^{\phi}=\prod_A H_A^{\varepsilon_A a_A\frac{D-2}{\Delta_A}},
\label{dilaton_sol} \ee
where $\Delta_A=(q_A+1)(D-q_A-3)+\frac{1}{2}a_A^2(D-2)$, 
$\varepsilon_A=+(-)$ if the corresponding brane is electrically
(magnetically) charged and $\delta^{(i)}_A=D-q_A-3$ or $-(q_A+1)$ depending
on whether the direction of $y_i$ is parallel or perpendicular to the
$q_A$-brane. Note that $\Delta_A=16$ and 18 for all the branes of,
respectively, 10 and 11 dimensional supergravities. To recapitulate,
in order to build up a metric according to the harmonic
superposition rule, we have to include a factor of 
$H_A^{-2\frac{D-q_A-3}{\Delta_A}}$ in front of each coordinate longitudinal
to the $q_A$-brane (including the time direction), and a factor of 
$H_A^{2\frac{q_A+1}{\Delta_A}}$ in front of each transverse coordinate,
and this has to be done for each brane in the configuration.

In the process of finding \rref{metric_sol} and \rref{dilaton_sol}, we 
did not use the ${R^a}_b$ off-diagonal components of the Einstein
equations. These have by now reduced to a set of algebraic conditions,
that for consistency impose the following pairwise intersection rule
for $\bar{q}=\mbox{dim}(\cap)$:
\be
\bar{q}+1=\frac{(q_A+1)(q_B+1)}{D-2}-\frac{1}{2}
\varepsilon_A a_A \varepsilon_B a_B. \label{intrule}
\ee

We now point out some remarks.

The formulae above \rref{metric_sol}, \rref{dilaton_sol} and
\rref{intrule} hold for $D-p>3$, in which case the space is
asymptotically flat, as well as for $D-p=2$ or 3, where the equation
\rref{multicenter} does not hold any more, i.e. the $H_A$'s do not
tend to a finite value at infinity. In that cases the solutions have to
be considered rather formally. Also for a Euclidean signature the
same formulae hold, without the obligation for the time coordinate to be
always longitudinal to all the branes (however the electric fields
have to be imaginary).

The total mass of these configuration is, as expected, the sum of the
masses of each constituent brane, which are equal to the charges:
$M=\sum M_A = \sum Q_A$.

All the solution above have a functional dependence restricted to the
overall transverse space. Some configurations which exist in string
theory, as the two D5-branes intersecting over a string (the
$\nu=8$ configurations in \cite{green_lect}) and their duals, are thus excluded
since they correspond to the solutions discussed in \cite{gkt}, where
the functions depend on the `relative transverse' coordinates.

As already stated above, we did not consider for simplicity
non-diagonal metrics. One can nevertheless find the solutions
involving KK travelling waves and KK monopoles applying some duality
transformation on the solutions above, since all KK charges are 
related by U-duality to the RR and NSNS charges. A classification
of the intersections involving also KK branes can be found in
\cite{bergshoeff}.

\section{Intersections in String and M-theory and Black Hole Entropy}
We can now specialize the formula above \rref{intrule} to the case
of $D=10$ and 11 maximal supergravities. 
Actually, this is done straightforwardly
specifying $D$ and the dilaton couplings $a_A$. For $D=11$ we
simply have $a=0$ since there is no dilaton. For $D=10$ IIA and IIB
theories, we have $a=-1$ for the NSNS 3-form field strength and
$\varepsilon a={1\over 2}(3-q)$ for a $q$-brane carrying RR electric
or magnetic charge.

As a first application, we will use the metric \rref{metric_sol} to
derive the number of charges one needs to build up a (supersymmetric)
extreme black hole with non-vanishing horizon area in a definite
number of non-compact dimensions. Let us define $\bar{D}=D-p$ 
($D=10,11$). Then in the Einstein frame the Bekenstein-Hawking 
entropy is proportional to the horizon area defined as follows:
\[ S\sim V(\mbox{compact space}) A(S^{D-p-2}) |_{r=0}. \]
Using \rref{metric_sol} in the case where all the harmonic functions
are centered at the same point $r=0$ (the horizon), we have:
\[
\begin{array}{rcl}
S&\sim& \prod_A H_A^{1/2} r^{\bar{D}-2} \\
& \sim & \prod_A Q_A^{1/2} r^{-{1\over 2}{\cal N}(\bar{D}-3)+\bar{D}-2}\\
&\sim & \prod_A Q_A^{1/2},
\end{array}\]
the last relation being true only provided the following relation
between ${\cal N}$ and $\bar{D}$ holds:
\be
{\cal N}=2{\bar{D}-2 \over \bar{D}-3}. \label{numcharges}
\ee
This relation has only two integer solutions, which are:
\be \bar{D}=5, \quad {\cal N}=3 \qquad \mbox{and} \qquad 
\bar{D}=4, \quad {\cal N}=4. \label{4and5dim} \ee
This also proves that there are no (stringy) extreme black holes
with non-zero entropy in $\bar{D}\geq 6$.

All the solution described in \rref{metric_sol}--\rref{intrule}
can be shown to be supersymmetric, and the fraction of preserved
supersymmetry is in general at least $1/2^{\cal N}$. For instance,
the ${\cal N}=3$ and ${\cal N}=4$ solutions discussed just above
both preserve $1/8$ of supersymmetry, thus providing an example
and a counter-example to the `$1/2^{\cal N}$ rule'.

We can now summarize all the possible pairwise intersections
between the branes which appear in string/M-theory. We use
the notation $q_A \cap q_B =\bar{q}$. This rules appeared case by case
in the literature, following from rather different arguments, in
\cite{strom_open,town_mbranes,tseytlin_noforce}.

In $D=11$, we have for the M-branes:
\be 2\cap2=0, \qquad 2\cap5=1, \qquad 5\cap5=3. \label{mbranes} \ee

In $D=10$, for the intersections between D-branes we have
generically $q_1\cap q_2={1\over2}(q_1+q_2-4)$, which gives the
following three cases:
\begin{eqnarray}
q\cap q &=& q-2 \label{samed}\\
(q-2)\cap q&=& q-3 \label{opend}\\
(q-4)\cap q&=& q-4 \label{withind}
\end{eqnarray}
The last case \rref{withind} can be interpreted as a D$(q-4)$-brane
within a D$q$-brane as in \cite{doug}. 

The intersections involving NSNS branes are:
\begin{eqnarray}
1_F\cap 5_S&=&1 \label{withinns}\\
1_F\cap q_D&=&0 \label{dbranes}\\
q_D \cap 5_S &=& q-1, \qquad 1\leq q\leq6 \label{ddbranes}
\end{eqnarray}
where the subscripts $F$, $S$ and $D$ denote respectively fundamental
strings, solitonic 5-branes and D-branes.

It is interesting to see how all these intersections come on an equal
footing in this framework, while they have a very different origin
in the underlying theories.

\section{When the Intersection is Actually a Boundary}
The second case in \rref{mbranes} and the cases \rref{opend}, 
\rref{dbranes} and \rref{ddbranes} all have a common feature: the
intersection has the same dimension as the (would-be) boundary of one of the
two branes. Are we allowed to consider
each $p-1$ dimensional intersection as the boundary of an open
$p$-brane tied to the world-volume of the other brane? 
The case \rref{dbranes} is effectively consistent with
the picture of fundamental strings ending on D-branes, but for the
other cases we do not have any quantum description of the phenomenon.

From the supergravity point of view, all these branes appear to be
closed. The intersection is not localized in the compact space, rather
all the branes are `smeared' over the transverse compact directions.
In this sense there is little distinction between closed branes
and open branes with both ends joined.

If we want to go deeper into the consideration of open branes, we
need some additional input.
For the opening of the branes to be consistent, we need
a conservation law for the charge carried by the open brane.
We will now see that such a conservation law exists, provided the
boundaries of the open brane are constrained to live on another
brane.

In words, the mechanism goes as follows. The charge carried by the 
brane is conserved when the brane is open if the boundary carries itself
a charge in the effective theory on the world-volume of the 
brane on which it is constrained. 
In this way, each brane which can act as a D-brane for other
branes (including strings) has an effective world-volume theory
whose field content is determined by this mechanism.

It has to be noted that branes ending on other branes were used
in \cite{hanawitt,elitzur} and all the following literature 
to study field theory phenomena, such as
dualities, from the dynamics of brane configurations.

There are two complementary approaches to the charge conservation. One is
due to Townsend \cite{surgery} and crucially makes use of the Chern-Simons
terms in the supergravity equations of motion. The other is based on the
gauge invariance of the open brane world-volume action. We work out here
both approaches for a definite example, a D2-brane ending on a D4-brane
in IIA string theory. The general case is described in \cite{aehw}.

The equation of motion for the 4-form field strength has to be supplemented
by the Chern-Simons term present in the full IIA supergravity action,
and by the source due to the presence of the D2-brane. The equation thus
reads, neglecting the dilaton and all numerical factors:
\be
d*F_4=F_4 \wedge H_3 +Q_2 \delta_7. \label{eqcssource}
\ee
Since there is also a D4-brane in the configuration, the Bianchi
identity for $F_4$ is also modified by a source term:
\be dF_4=Q_4 \delta_5. \label{bisource} \ee
On the other hand, due to the absence of NS5-branes, $H_3$ can be
globally defined as $H_3=dB_2$. The equation \rref{eqcssource} can be
rewritten as:
\be 
d(*F_4 -  F_4 \wedge B_2)= Q_2 \delta_7 + Q_4 \delta_5 \wedge B_2.
\label{eqcssource2} \ee
We can now integrate both sides of this equation over a 7-sphere $S^7$
which intersects the D2-brane only once (this is possible only if the
D2-brane is open). The result is:
\be 0=Q_2+Q_4 \int_{S^2} \hat{B}_2, \label{eqcsint} \ee
where the hat denotes the pull-back to the world-volume of the D4-brane
of a space-time field.

We see that the Chern-Simons term indicates the presence on the world
volume of the D4-brane of a 2-form field strength, for which the
(string-like) boundaries of the D2-branes act as magnetic charges.
As we will discuss shortly, the gauge invariant combination is
${\cal F}_2=dV_1-\hat{B}_2$. The presence of the Chern-Simons term
in \rref{eqcssource} ensures the consistency between charge conservation
of the open D2-brane and gauge invariance of the world-sheet action of
the open fundamental 
string. It is essential to note that a CS term exists for each case
mentioned above which could lead to the opening of one brane.

Considering now the world-volume action of the D2-brane,
we know that there is a minimal coupling to the RR 3-form potential:
\[ I_{D2}=Q_2 \int_{W_3} \hat{A}_3 +\dots \]
When the D2-brane is open, the gauge transformation $\delta A_3=d\Lambda_2$
becomes anomalous:
\[ \delta I_{D2}=Q_2 \int_{(\partial W)_2} \hat{\Lambda}_2. \]
The standard way to cancel this anomaly is by constraining the
boundary $(\partial W)_2$ to lie on the D4-brane world-volume
where a 2-form gauge potential $V_2$, transforming as $\delta V_2=
\hat{\Lambda}_2$, couples to it. The boundary of the D2-brane is now
an electric source for the 3-form field strength built out from
this potential. Again, the gauge invariant combination is given by
${\cal G}_3=dV_2-\hat{A}_3$.

The analysis of the Goldstone modes of broken supersymmetry and of
broken translation invariance, and the requirement that these bosonic
and fermionic modes still fit into a representation of the unbroken 
supersymmetries, forces us to identify the two field strengths
${\cal F}_2$ and ${\cal G}_3$ by an electric-magnetic duality on the
D4-brane world-volume:
\[ {\cal F}_2=\star {\cal G}_3. \]

Moreover, we could have analyzed instead the (more familiar) configuration
of a fundamental string ending on the D4-brane. We would have found
that its end point behaves like an electric charge for the 2-form
field strength and like a magnetic charge for the 3-form field
strength. Thus we conclude that the boundaries of the
string and of the membrane are electric-magnetic dual objects on the
world-volume of the D4-brane.

Let us now review the outcome of the analysis above when applied to
all the cases discussed at the beginning of this section.

\begin{itemize}
\item
All the D$p$-branes have a world-volume effective theory which can
be formulated in terms of a 2-form field
strength ${\cal F}_2$. The electric charges are the end points of the
fundamental strings, while the magnetic charges are the boundaries
of the D$(p-2)$-branes (as in \cite{strom_open}). Note the interesting case
of the D3-brane on the world-volume of which the S-duality between
fundamental strings and D-strings becomes electric-magnetic duality
between their end points. The presence of the 2-form field strength
is in this case supported by the quantum stringy computation
which gives super-Yang-Mills as the low energy effective action of
the D-branes.
\item
On the world-volume of the IIA NS5-brane, we can have the boundaries
of the D2- and D4-brane. The boundary of the D2-brane is self-dual
and thus couples to a self dual 3-form field-strength, while
the boundary of the D4-brane couples magnetically to a scalar potential.
This scalar potential is nothing else than the 11th direction which
remains after reduction of the M5-brane action. There is also a limiting
case here: from \rref{ddbranes} we can see that a D6-brane can end on
a NS5-brane. In this case however the NS5-brane {\em is} the boundary
of the D6-brane, much in the same way as the D0-branes are the
end points of fundamental strings. The charge conservation in these cases
has to be treated in a somewhat different way, e.g. one needs
the presence of D8-branes resulting in a non-vanishing cosmological
constant, see \cite{massive}.
\item
For the IIB NS5-brane, again all the IIB D-branes can have boundaries
on it. The D1- and the D3-brane boundaries are respectively the electric
and magnetic charge related to a 2-form field strength 
$\tilde{\cal F}_2=d\tilde{V}_1-\hat{A}_2^{RR}$
which can be considered the S-dual of the ${\cal F}_2$ field on the
D5-brane. The boundary of the D5-brane couples electrically to a 
(non-propagating) 6-form field strength ${\cal G}_6$. This 6-form
should be related to the mass term in the IIB NS5-brane action as discussed
in \cite{town_sl2z}, and could play a role in the definition of 
an $SL(2,Z)$ invariant IIB 5-brane action. Indeed, by S-duality we
should also have the possibility of a NS5-brane ending on a D5-brane.
\end{itemize}

Let us conclude this contribution with some speculations about the emission
of closed branes. The idea is to revert to argument which leads, from
the intersecting configurations, to the open brane configurations.

Suppose that we have an open $q$-brane with both boundaries tied to the same
closed $p$-brane. The world-volume of the open $q$-brane wraps some transverse
compact directions in order to have a definite space-time charge.
The two boundaries are, from the point of view of the world-volume of
the closed $p$-brane, two opposite charges. This has as a consequence that
the configuration is not BPS, and therefore not supersymmetric, 
since there is a force between these two charges. Now the two opposite
charges can meet and annihilate, which from the $q$-brane point of
view means that the two boundaries meet and reconnect. The $q$-brane
is now closed, and moreover the bound state it constitutes with the 
closed $p$-brane has vanishing binding energy. Nothing then prevents it 
to leave the $p$-brane (there is some energy left due to the attraction
of the two opposite charges). This is thus a very rough picture of how
a $p$-brane could emit other closed branes.

This mechanism has been shown for the emission of closed fundamental 
strings by D-branes \cite{emission}. To describe in a more detailed way
the emission of higher branes, a quantum theory of the latter is still
lacking. Matrix theory \cite{bfss} could be a suitable framework to
treat this problem (see e.g. \cite{openmatrix} for a proposal on
open membranes in Matrix theory).

The interest to study these processes is certainly very high. Brane
emission could be the dominant process when Hawking radiation is
considered at strong coupling, $g_s\gg 1$, or for the emission from
black NS 5-branes.

\acknowledgements
I would like to thank my collaborators in the work which was the subject
of this talk, F.~Englert, L.~Houart and P.~Windey, and the organizers
of the Carg\`ese summer school for giving me the opportunity to present
this work. The author is a research assistant (Aspirant) of the
Fonds National de la Recherche Scientifique (Belgium).

\end{document}